\documentclass{PoS}

\newcommand{\pT}[0]{p_{\mathrm{T}}}						
\newcommand{\mpT}[0]{\left < p_{\mathrm{T}} \right >} 							
 							
\newcommand{\moment}[0]{\left < p_{\mathrm{T}}^n \right >} 							

\newcommand{\nTr}[1]{n_{\mathrm{par}}(#1)} 							
 							
\newcommand{\nEv}[1]{n_{\mathrm{evt}}(#1)} 							
\newcommand{\nEvHat}[1]{\hat{n}_{\mathrm{evt}}(#1)}

\newcommand{\Nacc}[0]{N_{\mathrm{acc}}} 							
\newcommand{\Nch}[0]{N_{\mathrm{ch}}}

\newcommand{\Nchprim}[0]{N_{\mathrm{ch}}^{'}} 							

\newcommand{\res}[0]{P(\Nacc|\Nch)} 							
 							
\newcommand{\unf}[0]{P(\Nch|\Nacc)}

\newcommand{\resprimch}[0]{P(\Nacc|\Nchprim)}

\newcommand{\abs}[1]{\left| #1 \right|} 								
\newcommand{\ddd}[3]{\frac{d^2 #1}{d #2\ d #3}} 							
\let\baraccent=\= 													
\renewcommand{\=}[1]{\stackrel{#1}{=}} 								

\title{Bayesian unfolding of charged particle $p_{\mathrm{T}} $ spectra with ALICE at the LHC}
\ShortTitle{Bayesian unfolding of charged particle $p_{\mathrm{T}} $ spectra with ALICE at the LHC}

\author{\speaker{Mario Kr\"uger} for the ALICE Collaboraion\\
        Institut f\"ur Kernphysik, Goethe-University Frankfurt\\
        E-mail: \email{mario.kruger@cern.ch}}

\abstract{The study of  the Quark-Gluon Plasma created in ultrarelativistic heavy-ion collisions at the CERN-LHC is complemented by reference measurements in  proton-lead (p--Pb) and proton-proton (pp) collisions, where the effects of multiple-parton interactions and hadronization beyond independent string fragmentation can be investigated. In these proceedings, we present a Bayesian unfolding procedure to reconstruct the correlation between transverse momentum ($ p_{\mathrm{T}}$) spectra of charged particles and the corresponding charged-particle multiplicities $N_{\mathrm{ch}}$. The unfolded spectra are presented in single multiplicity ($\Delta N_{\mathrm{ch}}$ = 1) bins and are used to derive moments of the $p_{\mathrm{T}} $ distributions. We illustrate the unfolding procedure of the $ p_{\mathrm{T}} $ spectra with a Monte Carlo simulation for pp collisions at a centre-of-mass energy of $\sqrt{\mathrm{s}}= 5.02$ TeV.}

\FullConference{XIII Quark Confinement and the Hadron Spectrum - Confinement2018\\31 July - 6 August 2018\\Maynooth University, Ireland}

\begin{document}

\section{Introduction}
\begin{figure}[t]
	\centering
	\includegraphics[width=0.49\textwidth]{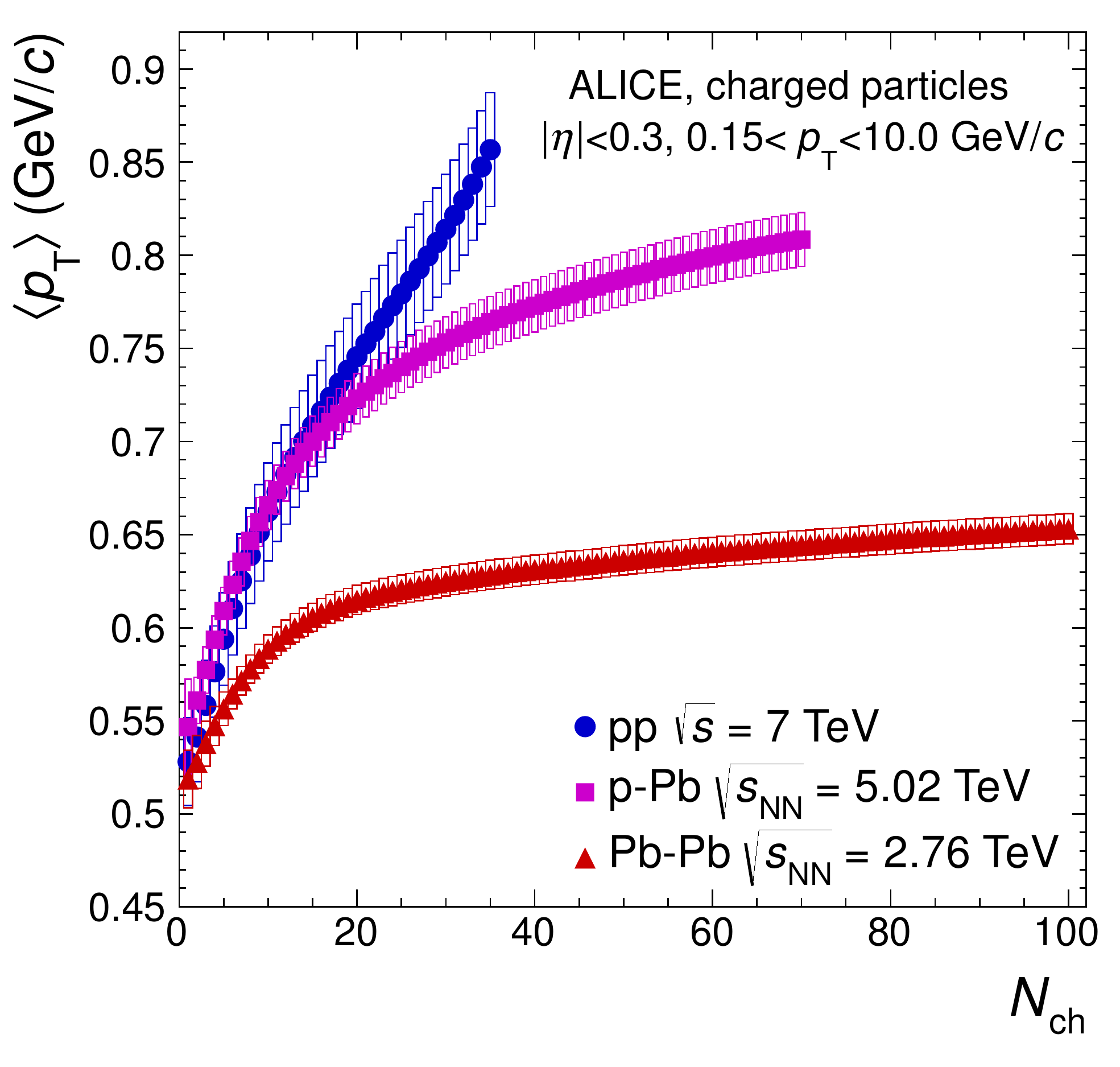} 
	\includegraphics[width=0.49\textwidth]{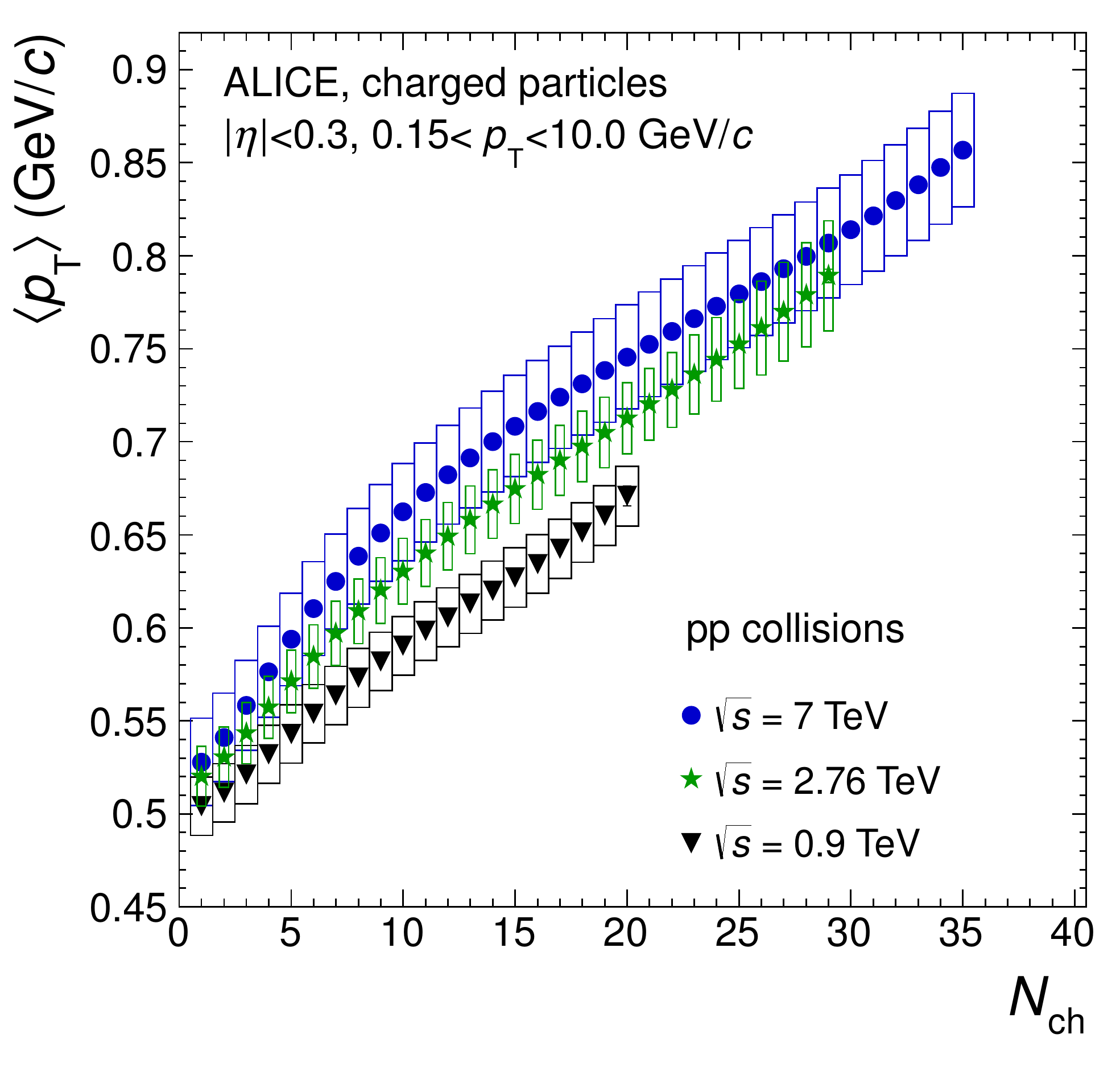} 
	\caption{The left panel shows $\mpT$ vs. $\Nch$ for pp, p--Pb and Pb--Pb collisions at the different centre-of-mass energies available at the time of publication.
			 On the right panel $\mpT$ vs. $\Nch$ is shown for pp collisions at $\sqrt{\mathrm{s}}= $\ 0.9 TeV, 2.76 TeV and 7 TeV \cite{meanpt-paper}.}
	\label{fig:paperPlots}
\end{figure}

High energy heavy-ion collisions at the Large Hadron Collider (LHC) provide the possibility to create a hot and dense deconfined state of matter in the laboratory.
One commonly used observable to study the properties of such medium is the transverse momentum ($\pT$) spectrum of the abundantly produced charged particles.
In central heavy-ion collisions, a suppression of the charged-particle yield with respect to the yield expected from a simple superposition of nucleon-nucleon collisions is observed, indicating a significant energy loss of partons traversing the quark gluon plasma.
For more peripheral collisions, it is possible to directly compare the $\pT$ spectra of heavy-ion and proton-nucleus as well as proton-proton collisions at the same final state charged-particle multiplicity $\Nch$.
It is very common to present the mean value $\mpT$ of these spectra as a function of the multiplicity, as shown in the left panel of figure \ref{fig:paperPlots} for an ALICE measurement of Pb--Pb, p--Pb and pp collisions at different centre-of-mass energies.
For all three collision systems an increase of $\mpT$ with $\Nch$ is observed, being strongest for proton-proton collisions.
In Pb--Pb, the average $\pT$ is significantly lower, which might be due to substantial rescattering of the particles produced in the collision.
The $\mpT$ vs. $\Nch$ observed in p--Pb shows a steep rise similar to the one in pp at low multiplicities and a Pb--Pb like saturating trend at high multiplicities.
With the more recent ALICE datasets it is now possible to perform this measurement at the same centre-of-mass energy per nucleon pair for pp, p--Pb and Pb--Pb, which will provide a better comparison of the three collision systems.
In addition, the variety of pp datasets measured with ALICE allows for an extensive energy scan of the multiplicity dependent $\mpT$, complementing the results shown in the right panel of figure~\ref{fig:paperPlots}.
These new measurements can be used to further investigate the underlying particle production mechanisms and provide more insight into the effects of multiple-parton interactions and hadronization beyond independent string fragmentation.
By extending the $\mpT$ vs. $\Nch$ observable to a precise measurement of the complete $\pT$ spectra as function of multiplicity, ALICE will be able to provide even more rigorous constraints for theoretical models.
\\
In these proceedings we present an analysis method based on a Bayesian unfolding technique, which is used to obtain the $\pT$ spectra as function of $\Nch$.
The procedure is illustrated by means of a Monte Carlo simulation of pp collisions at $\sqrt{\mathrm{s}}= 5.02$ TeV using the PYTHIA8 \cite{pythia} event generator and a virtual model of the ALICE detector.

\section{Analysis}
\begin{figure}[t]
	\centering
	\includegraphics[width=0.49\textwidth]{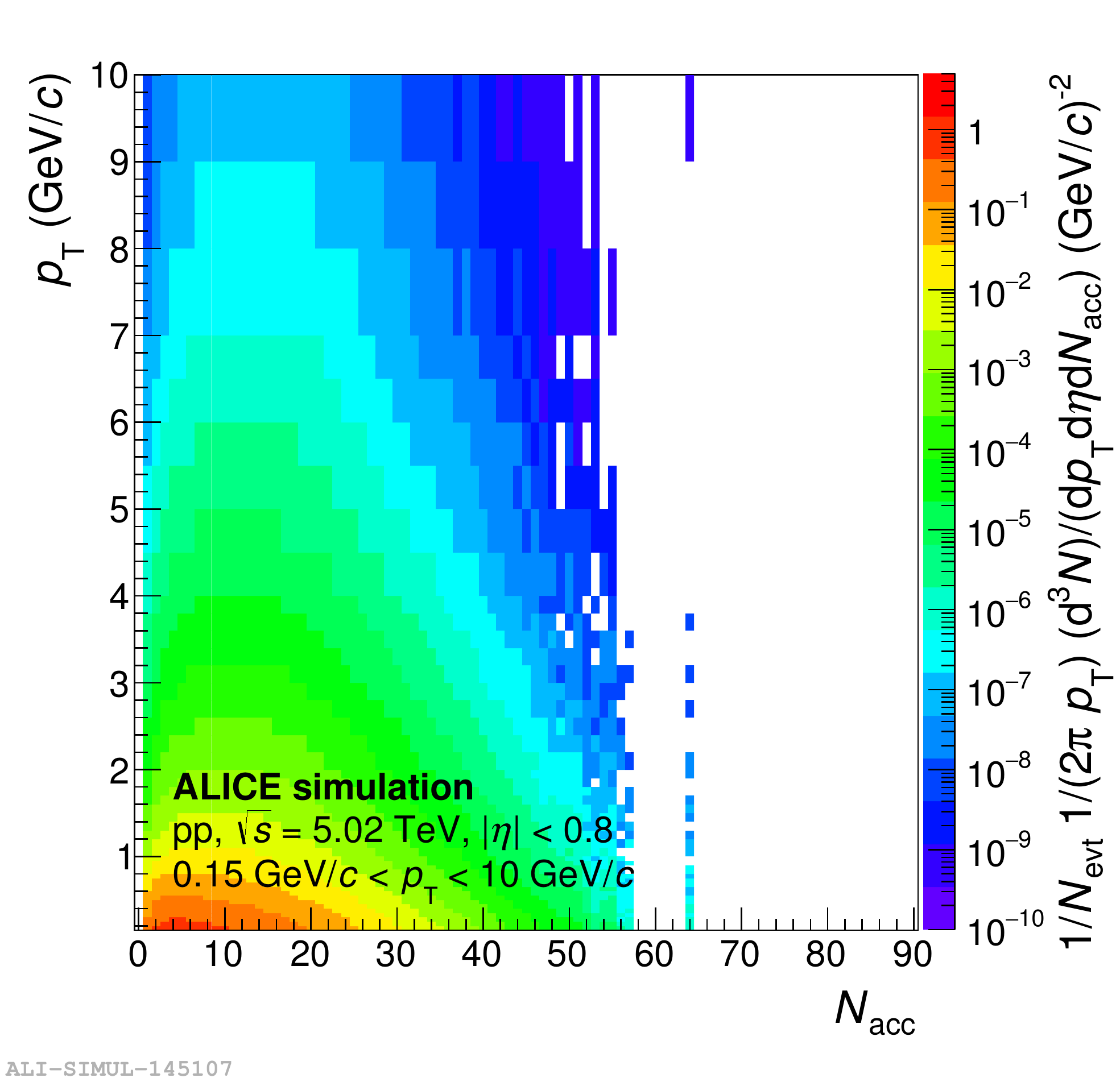} 
	\includegraphics[width=0.49\textwidth]{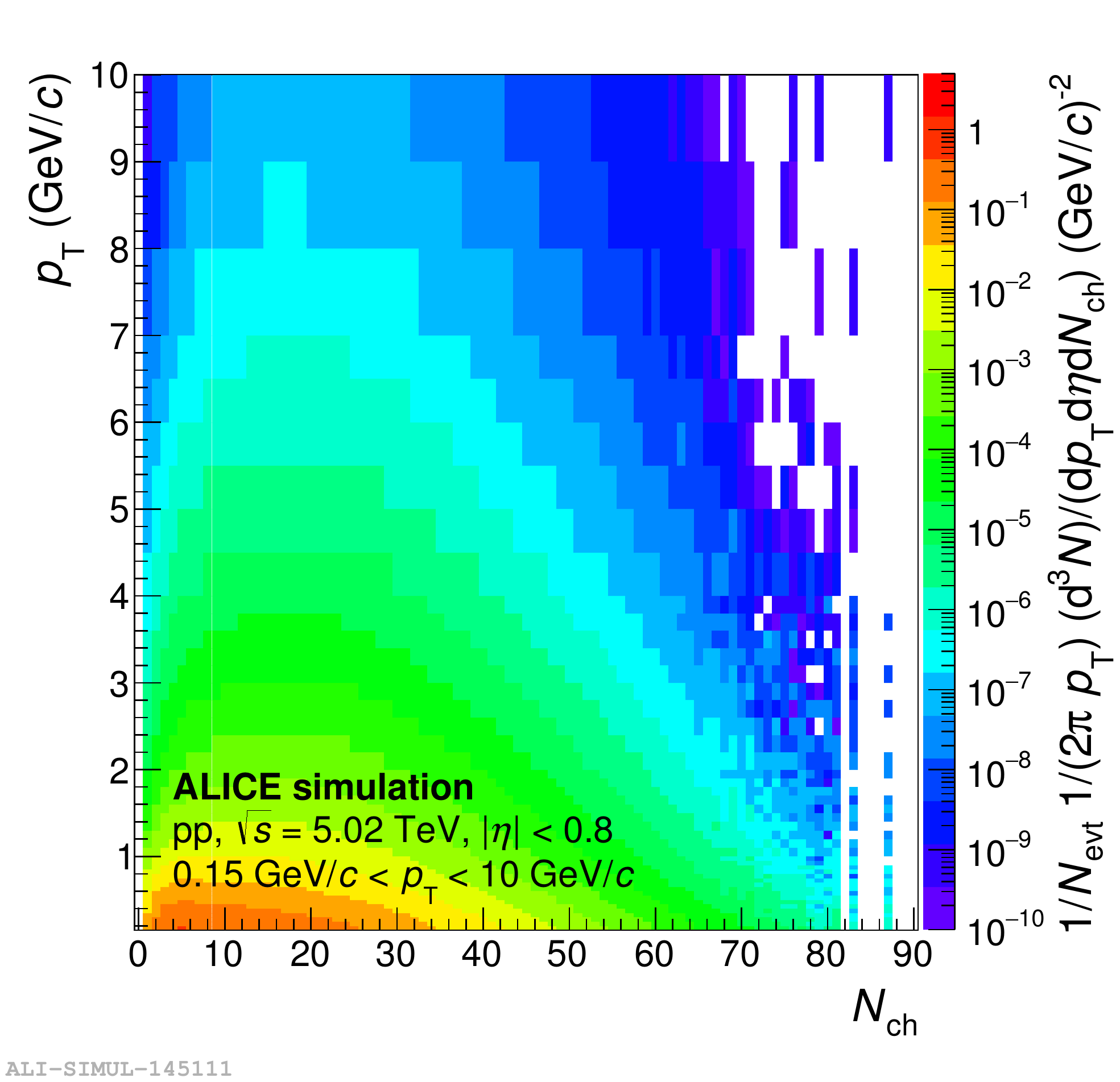} 
	\caption{Transverse momentum spectra from an ALICE simulation of pp collisions at $\sqrt{s} = 5.02\ \mathrm{TeV}$ as a function of the measured multiplicity $\Nacc$ (left) and as a function of the true multiplicity $\Nch$  (right) after unfolding.}
	\label{fig:multPt}
\end{figure}

\begin{figure}[t]
	\centering
	\includegraphics[width=0.49\textwidth]{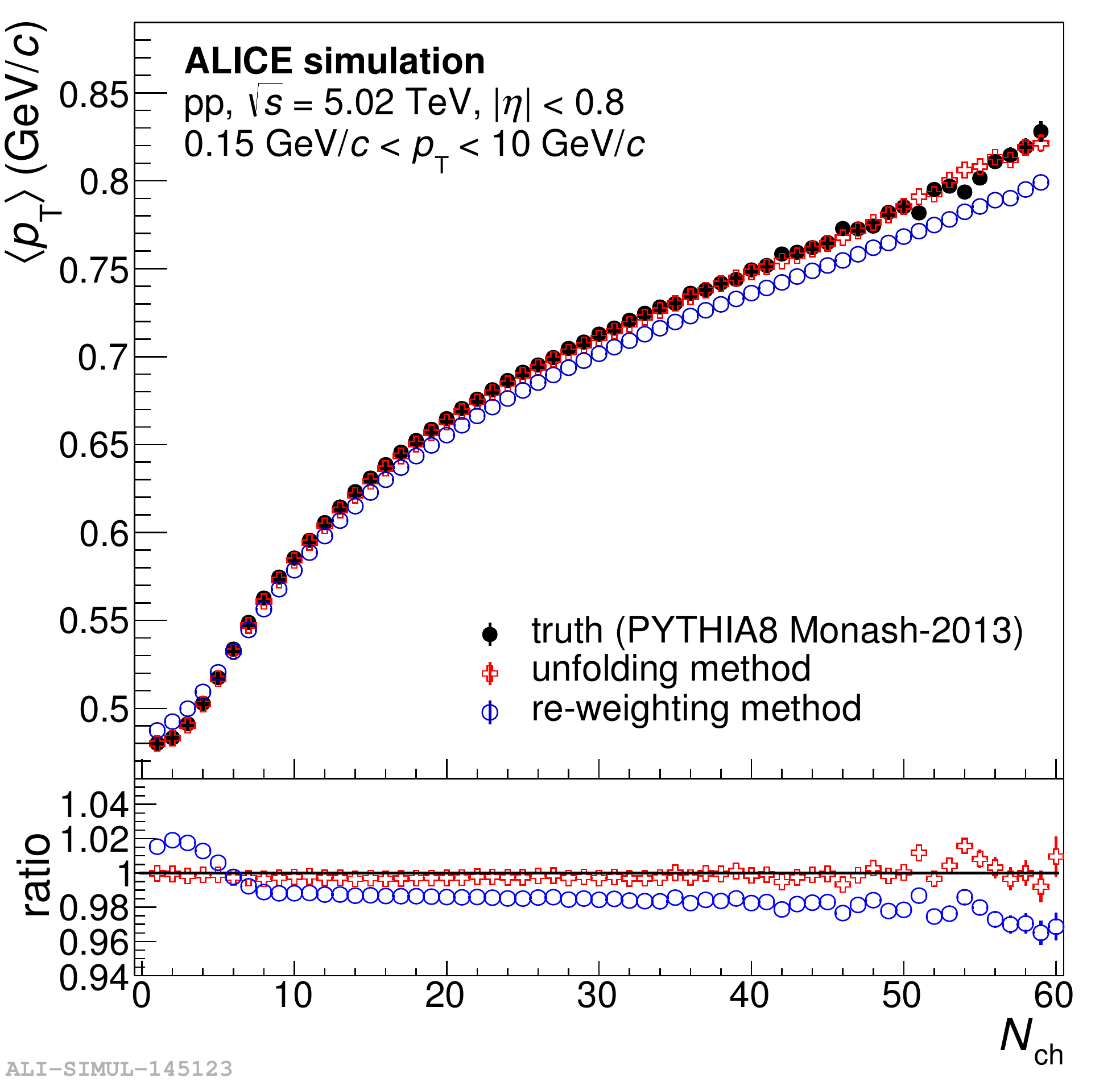} 
	\includegraphics[width=0.49\textwidth]{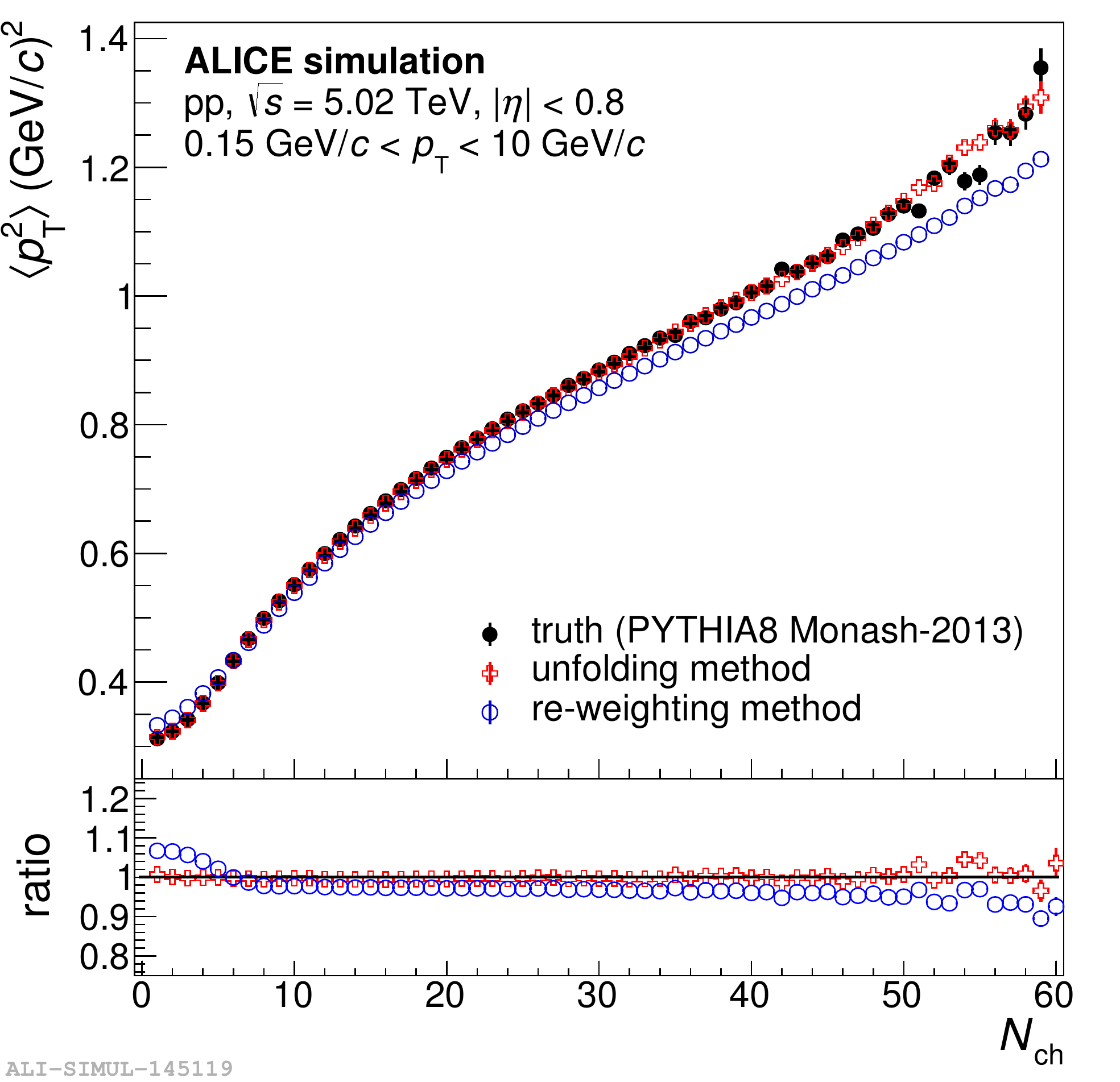} 
	\includegraphics[width=0.49\textwidth]{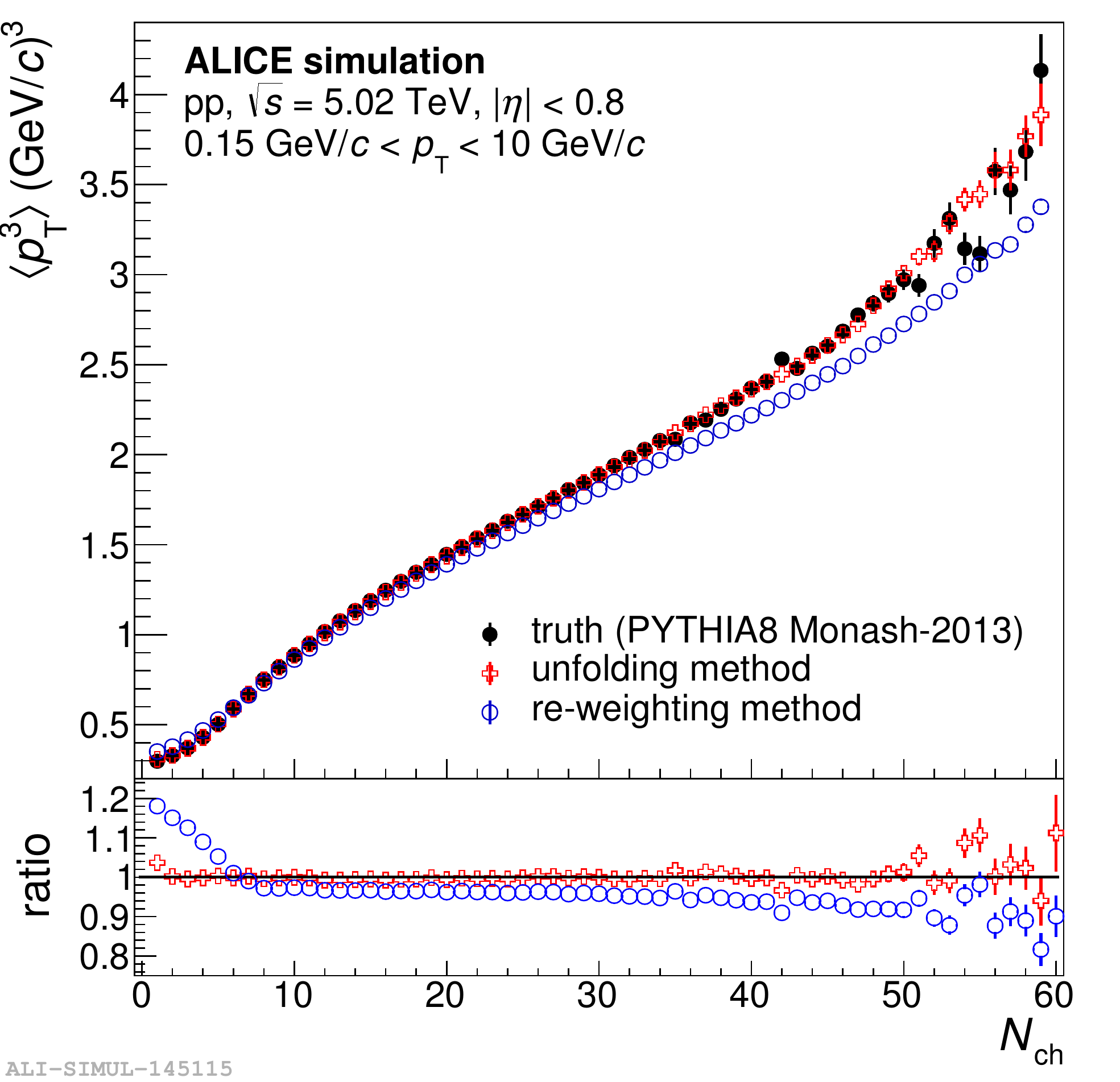} 
	\caption{Closure tests for the first three moments of the unfolded $\pT$ spectra from an ALICE simulation compared to corresponding closure tests of the re-weighting method. }
	\label{fig:closureTests}
\end{figure}

The ALICE experiment is very well suited to measure precisely the transverse momenta of charged particles.
Its large solenoid magnet creates a field bending the charged-particle trajectories, which are reconstructed using space point measurements in the Inner Tracking System (ITS) and the Time Projection Chamber (TPC).
A detailed description of the ALICE apparatus and its detector systems can be found in \cite{alice}.
All collision events considered in this analysis are required to fulfil the ALICE minimum-bias trigger condition.
The kinematic range of the charged-particle measurement is restricted to $ 0.15\ \mathrm{GeV}/c < \pT < 10\ \mathrm{GeV}/c $ and $ \abs{\eta}  < 0.8 $.
In order to ensure uniform acceptance and efficiency in this pseudorapidity window for all the detectors involved in the track reconstruction, the vertex is constrained to a maximal difference with respect to the nominal interaction point in the beam direction of $ \abs{V_z} < 10\ \mathrm{cm} $.
Only tracks fulfilling the strict quality criteria described in \cite{raa-paper} contribute to the measurement, which is then corrected for acceptance, efficiency and secondary contamination using a Monte Carlo simulation which incorporates a GEANT3 \cite{geant} model of the ALICE detector.
The resulting transverse momentum spectra are obtained as a function of the number of measured tracks $\Nacc$.
Due to the limited detector efficiency, this is not equal to the true number of charged particles produced in the collisions.
The probability to measure only $\Nacc$ of the $\Nch$ original particles is given by the detector response $\res$.
Events with a true multiplicity $\Nch$ can potentially be reconstructed with many different $\Nacc$ and therefore contribute to various measured multiplicity-dependent $\pT$ spectra.
Thus, the actual correlation between the $\pT$ spectra and their corresponding true multiplicities is smeared in the experiment.
\\
One approach to account for this smearing is to re-weight the moments $\moment(\Nacc)$ of the measured $\pT$  spectra with the detector response:
\begin{equation}
\moment(\Nch) = \sum_{\Nacc} \res \cdot \moment(\Nacc) \ .
\label{eq: re-weighting}
\end{equation}
This procedure was used to obtain the average transverse momentum as a function of $\Nch$ in multiple ALICE publications (eg. \cite{meanpt-paper} and  \cite{meanpt-paper-900}).
In contrast, the goal of this analysis is to perform an unfolding of the full spectral shape and obtain the entire $\pT$ spectra as a function of $\Nch$:
\begin{equation}
\frac{1}{N_{evt}} \frac{1}{2\pi \pT} \ddd{N}{\eta}{\pT}(\Nacc) \qquad\Longrightarrow\qquad \frac{1}{N_{evt}} \frac{1}{2\pi \pT} \ddd{N}{\eta}{\pT}(\Nch)  \ .
\end{equation}
This can be realized by means of an iterative unfolding procedure published by G. D'Agostini in 1995 \cite{agostini-bayes}, which is based on the Bayesian theorem \cite{bayes-original}. 
In the following, the corresponding algorithm is briefly sketched and formulated with the terminology of true and measured multiplicities.
\begin{figure}[t]
	\centering
	\includegraphics[width=1.0\textwidth]{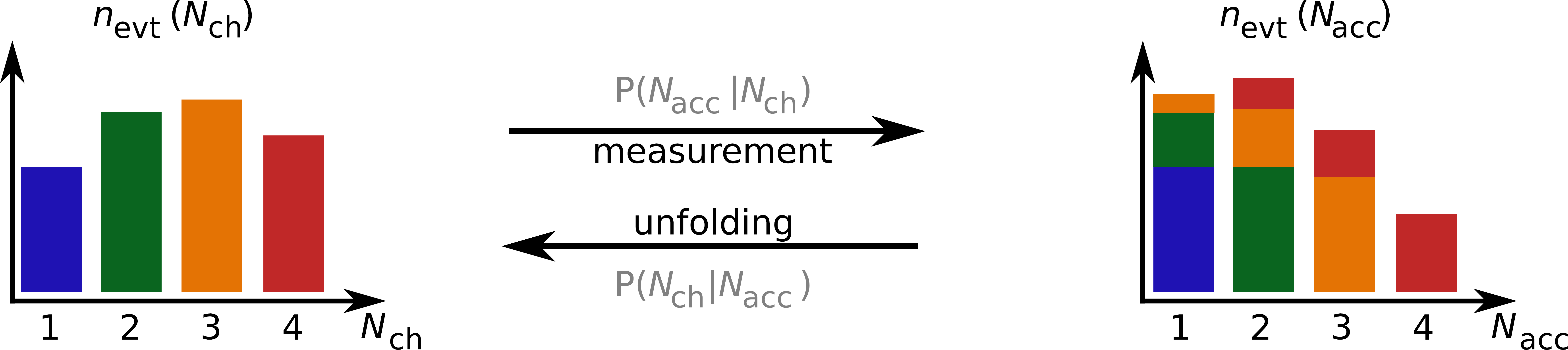} 
	\caption{Sketch illustrating the unfolding of multiplicity distributions.}
	\label{fig:illustrationUnfolding}
\end{figure}
\\
As illustrated in figure \ref{fig:illustrationUnfolding}, the multiplicity distribution of measured events $ \nEv{\Nacc} $ is a convolution of the true multiplicity distribution $ \nEv{\Nch} $  and the detector response $\res$.
For unfolding the measurement, one needs to find out how $ \nEv{\Nacc} $ is composed with respect to the various true multiplicities $ \Nch $.
The fraction of events which are measured with multiplicity $ \Nacc $, but originally have a true multiplicity $ \Nch $, is represented by the conditional probability $ \unf $.
Following D'Agostini this can be expressed as:
\begin{equation}
\unf = \frac{\res \cdot P(\Nch)}{\sum_{\Nchprim} \resprimch P(\Nchprim)} \ .
\label{eq: unf}
\end{equation} 
With an arbitrary choice for the  unknown probability distribution of the true multiplicities $ P(\Nch)$, a first guess for $ \unf $ can be calculated and subsequently used to obtain an estimate for the unfolded multiplicity distribution:
\begin{equation}
\nEvHat{\Nch} = \sum_{\Nacc} \unf\ \nEv{\Nacc} \ .
\label{eq: unfoldingProcedure}
\end{equation}
The $ P(\Nch)$ which can be inferred from this unfolded multiplicity distribution is more accurate than the arbitrary initial guess used in the beginning, since it is now constrained by the measurement of $ \nEv{\Nacc} $.
This suggests to update $\unf$ accordingly and repeat the process several times in order to obtain a good estimate for the true multiplicity distribution $\nEv{\Nch} $.
Further details on the iterative D'Agostini method can be found for example in \cite{agostini-bayes} and \cite{rooUnfold}.
\\
To make use of this procedure for the unfolding of multiplicity-dependent $\pT$ spectra, it is important to note that one is dealing with particle distributions $ \nTr{\Nacc} $ rather than event distributions $ \nEv{\Nacc} $.
This implies, that the detector response needs to be defined on particle level as well, representing the probability of a charged particle originating from an event with multiplicity $\Nch$ to be measured in an event with multiplicity $\Nacc$.
The unfolding is performed differentially in $\pT$, i.e. the iterative D'Agostini procedure is applied separately for all $\pT$ intervals of the measurement.
\\
Figure \ref{fig:multPt} presents the outcome of a PYTHIA8 Monte Carlo simulation of pp collisions at a centre-of-mass energy of $\sqrt{\mathrm{s}}= 5.02$ TeV  propagated through a virtual model of the ALICE detector.
The left panel depicts the corresponding measured $\pT$ spectra of primary charged particles in a kinematic range of $ 0.15\ \mathrm{GeV}/c < \pT < 10\ \mathrm{GeV}/c $ and $ \abs{\eta}  < 0.8 $ as a function of the measured multiplicity $\Nacc$.  The right panel shows these spectra as a function of $\Nch$ after applying the $\pT$-differential unfolding.
As a self consistency check (closure test) of the unfolding method, these resulting spectra are compared to the multiplicity-dependent $\pT$ spectra produced by the underlying PYTHIA event generator (MC truth information).
Figure \ref{fig:closureTests} shows the first three moments of the deconvoluted  $\pT$ spectra in red markers and the moments of the true spectra as black markers.
Additionally, the moments are calculated with the re-weighting method described in equation \ref{eq: re-weighting} and plotted here in blue markers.
The bottom panels show the ratios of the results from both the unfolding and the re-weighting method with respect to the MC truth information.
Evidently, the unfolding yields very accurate results for all three moments of the spectra. 
Thus, we conclude that the iterative D'Agostini method is very successful in unsmearing the detector effects over the whole range of multiplicities and in good approximation recovers the full spectral shape.
The re-weighting method on the other hand apparently introduces a bias of the order of a few percent, which becomes worse for the higher moments.
In previous publications of $\mpT$ vs. $\Nch$ this non-closure contributed to the assigned systematic uncertainties.
Therefore, applying the unfolding procedure described in these proceeding to real ALICE data will result in a more precise measurement of this observable.	
	

\section{Summary and outlook}
The correlation between charged-particle $\pT$ spectra and multiplicity is a good observable to study charged-particle production mechanisms.
It poses a challenge to reconstruct this correlation from experimental data, since the multiplicity measurement is smeared as a result of limited detector efficiencies.
This problem can be tackled by means of an unfolding procedure based on the  iterative D'Agostini method (often also called 'Bayesian unfolding').
In these proceedings we present the results of this approach for the example of an ALICE simulation of pp collisions at $\sqrt{s} = 5.02\ \mathrm{TeV}$.
By means of a Monte Carlo closure test we show that this method significantly improves the $\mpT$ vs. $\Nch$  measurement with respect to a previously used procedure.
With this new tool set at hand, it is now possible to precisely determine the multiplicity dependent charged-particle $\pT$ spectra for the variety of ALICE datasets available.
The resulting energy and system size dependent measurements can provide an important benchmark for modern event generators.

\bibliographystyle{JHEP}
\bibliography{Sources}

\end{document}